\documentclass[onecolumn]{webofc}
\usepackage[utf8]{inputenc}
\usepackage{hyperref}
\usepackage{tikz}
\usepackage[varg]{txfonts}   
\newcommand{\trento}{\textsc{T$_\text{R}$ENTo}}
\newcommand{\as}{\alpha_\mathrm{s}}
\newcommand{\xpom}{x_\mathbb{P}}
\usepackage{wrapfig}
\begin{document}
\title{Theoretical developments on the initial state in relativistic particle collisions}
%
%

\author{\firstname{Heikki} \lastname{M\"antysaari}\inst{1,2}\fnsep\thanks{\email{heikki.mantysaari@jyu.fi}} 
}

\institute{Department of Physics, University of Jyväskylä,  P.O. Box 35, 40014 University of Jyväskylä, Finland
\and
           Helsinki Institute of Physics, P.O. Box 64, 00014 University of Helsinki, Finland           
          }

\abstract{%
  We discuss recent progress towards developing accurate initial state descriptions for heavy ion collisions focusing on weak coupling based approaches, that enable one to constrain the high-energy structure of nuclei from  deep inelastic scattering or proton-nucleus collisions.  We review recent developments to determine the event-by-event fluctuating nuclear geometry, to describe gluon saturation phenomena at next-to-leading order accuracy, and to include longitudinal dynamics to the initial state descriptions.

  }
\maketitle
\section{Introduction}
\label{intro}
A crucial ingredient needed to simulate the space-time evolution in heavy ion collisions is the structure of the colliding nuclei at small momentum fraction $x$, which is the region probed in high-energy collisions.
A consistent description of the heavy ion collision initial state  together with e.g.  deep inelastic scattering (DIS) data has been achieved 
in approaches based on collinear factorization and Color Glass Condensate (CGC). In the EKRT model based on collinear factorization~\cite{Paatelainen:2012at} the partonic content of the nuclei is described in terms of nuclear parton distribution functions. 
In the CGC approach (implemented e.g. in the IP-Glasma~\cite{Schenke:2012wb} framework)  the DIS and p+A cross sections and the time evolution of the color fields immediately after the heavy ion collision are described in terms of the universal Wilson line correlators. 

There are also many other approaches to describe the initial state in heavy ion collisions including, for example, parametrization-based models such as \trento~\cite{Soeder:2023vdn} and different event generators (Pythia/Angantyr, EPOS, HIJING). In this contribution we, however, focus on weak coupling approaches with a direct connection to DIS and p+A collisions.

\section{Probing nuclear geometry in photon-nucleus scattering}

In heavy ion collisions the initial state eccentricities are transformed into momentum space anisotropies by the hydrodynamically evolving QGP. As such, a crucial input to the QGP simulations is the spatial distribution of nuclear matter at the initial condition and immediately after the collision before an approximatively thermalized QGP is formed.  As such, there has been extensive activity in recent years to constrain the event-by-event fluctuating shape for of the proton and heavy nuclei.

Exclusive processes where the total momentum transfer is the Fourier conjugate to the impact parameter directly probe the nuclear geometry~\cite{Klein:2019qfb}. Exclusive vector meson production and DVCS were extensively studied ad HERA. In recent years, vector-meson photoproduction has been studied in ultra peripheral photon-mediated collisions at RHIC and at the LHC where the photon-nucleus processes are available before the EIC era. Such measurements directly probe the nuclear geometry, down to very small $\xpom\sim 10^{-5}$, and are sensitive probes of non-linear QCD dynamics. Furthermore the potential to constrain nuclear PDFs in the poorly constrained small-$x$ region has also been investigated recently~\cite{Eskola:2023oos}.

When exclusive vector meson production has been measured as a function of momentum transfer~\cite{ALICE:2021tyx,LHCb:2022ahs}, a spectrum that is more steeply falling compared to the one obtained as a Fourier transform of the Woods-Saxon density profile is obtained. This can be interpreted as a signature of saturation phenomena that effectively transform the nuclear density profile towards the black disc shape at small $\xpom$~\cite{Mantysaari:2022sux}.

\begin{wrapfigure}{l}{0.4\textwidth}
 \centering 
\includegraphics[width=0.4\textwidth]{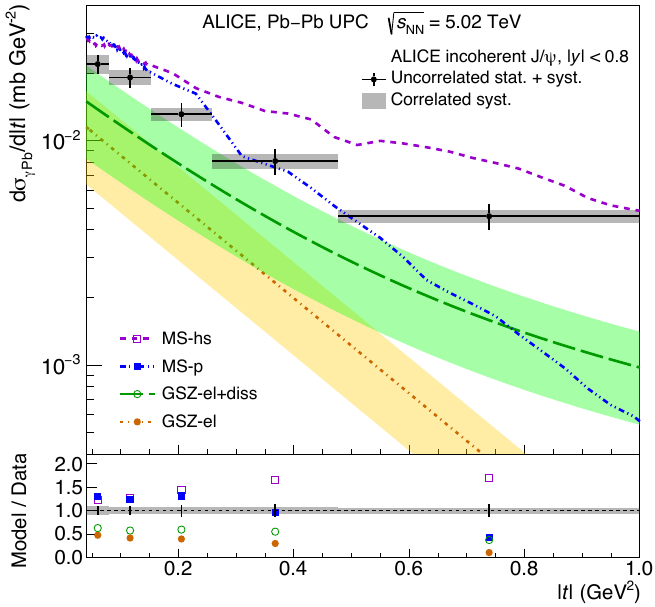}
\caption{Incoherent $\mathrm{J}/\psi$ production as a function of squared momentum transfer $|t|$ measured by ALICE compared to theory calculations with (MS-hs) and without (MS-p) nucleon substructure fluctuations. Figure from Ref.~\cite{ALICE:2023gcs}}
\label{fig:alice_incoh_t}
\quad
 \begin{minipage}{.49\textwidth}
 \end{minipage}
\end{wrapfigure}

In this conference, for the first time measurements of the incoherent $\mathrm{J}/\psi$ photoproduction cross section in ultra peripheral heavy ion collisions (i.e. in photon-nucleus collisions) as a function of the momentum transfer was reported by the ALICE and STAR collaborations~\cite{ALICE:2023gcs}. The incoherent production where the target dissociates is interesting, as it probes the event-by-event fluctuations in the target geometry~\cite{Mantysaari:2020axf}. For example, the HERA data has been shown to prefer significant event-by-event geometry fluctuations for the proton~\cite{Mantysaari:2016ykx}. The new ALICE data  is shown in Fig.~\ref{fig:alice_incoh_t}, where it is compared to CGC calculations that either use spherical nucleons or include an event-by-event fluctuating nucleon geometry constrained by the HERA data~\cite{Mantysaari:2017dwh}. Although the overall cross section is overestimated by the theory calculation (i.e. not enough nuclear suppression is obtained), the $t$-slope can be interpreted to prefer the calculation with nucleon substructure similar to that in protons at HERA kinematics. In addition to nucleon substructure fluctuations, in this conference recent progress towards probing the deformed structure of e.g. Uranium and Xenon in deep inelastic scattering  was presented~\cite{Mantysaari:2023qsq}.

\section{Gluon saturation at the precision level}
\label{sec:saturation}
Despite the fact that the leading order CGC calculations (that resum $\as \ln 1/x$ contributions to all orders) have been successful in describing large amount of small-$x$  data, it is crucial to develop the theory to the next-to-leading order accuracy to enable precision level comparisons with the current and future measurements. Over the last couple of years, there has been an extensive effort in the community to bring the theory calculations describing the gluon saturation phenomena to the next-to-leading order accuracy.

At small-$x$  cross sectios factorize to a convolution of Wilson lines and a hard impact factor. The energy dependence of the Wilson lines is described by perturbative Balitsky-Kovchegov or JIMWLK evolution equations. For a fully consistent NLO calculation, all these ingredients, including the non-perturbative initial condition for the high-energy evolution, need to be promoted to this order in $\as$. 

The NLO BK evolution equation became available already in 2007~\cite{Balitsky:2008zza}, and later the NLO JIMWLK equation has also been obtained~\cite{Balitsky:2013fea,Kovner:2013ona} although for that there is currently no known method to solve it numerically. Typically the non-perturbative initial condition describing the proton structure at moderately small $x$ has been extracted from fits to the proton structure function data~\cite{H1:2015ubc}. This became possible at NLO accuracy once the hard impact factor for DIS (the photon light front wave function at NLO) became available~\cite{Beuf:2017bpd} (see also Ref.~\cite{Dumitru:2023sjd} for a complementary approach based on proton valence quark wave function). The first NLO fit has been reported in Ref.~\cite{Beuf:2020dxl}, and recently also a successful description of both the total and heavy quark production cross sections in DIS has been obtained~\cite{Hanninen:2022gje}.

\begin{figure}[tb]
 \begin{minipage}{.49\textwidth}
 \centering 
\includegraphics[width=0.8\textwidth]{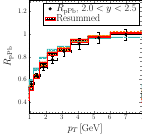}
\caption{Nuclear suppression factor for inclusive charged particle production in proton-lead collisions at the LHC compared to the LHCb data~\cite{LHCb:2021vww}. Figure from Ref.~\cite{Shi:2021hwx}.}
\label{fig:rpa_sinc}       
\end{minipage}
\quad
 \begin{minipage}{.49\textwidth}
 \centering
 \includegraphics[width=\textwidth]{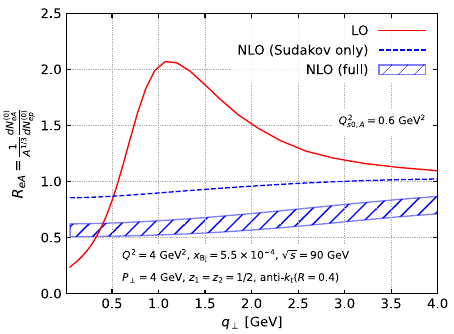}
 \caption{Nuclear suppression factor for dijet production in deep inelastic scattering in EIC kinematics as a function of dijet momentum $q_T$. Figure from Ref.~\cite{Caucal:2023fsf}.}
 \label{fig:rea_jet}
 \end{minipage}
\end{figure}

In addition to total DIS cross sections, impact factors for many other scattering processes are currently known at NLO. These include, for example, exclusive vector meson production~\cite{Mantysaari:2022kdm,Mantysaari:2022bsp,Mantysaari:2021ryb,Boussarie:2016bkq} and dijet/dihadron production~\cite{Caucal:2023fsf,Bergabo:2023wed,Taels:2022tza,Taels:2022tza} in DIS and inclusive hadron production in proton-nucleus collisions~\cite{Chirilli:2012jd,Stasto:2013cha,Ducloue:2017dit}. Recently, first phenomenological applications at NLO accuracy have also become available. In particular consistent NLO calculations (with the caveat that the NLO BK evolution equation is approximated by a leading order equation into which dominant higher order corrections have been resummed) compared to available data exist for exclusive light and heavy vector meson production~\cite{Mantysaari:2022kdm,Mantysaari:2022bsp,Mantysaari:2021ryb},  inclusive $\pi^0$ production in proton-lead collisions~\cite{Mantysaari:2023vfh} and dijet production in DIS~\cite{Caucal:2023fsf} (although in that case the initial condition for the small-$x$ evolution is not constrained by other collider data). Additionally, numerical results where leading order Wilson line correlators are used together with  NLO impact factors exist for charged hadron production in proton-nucleus collisions~\cite{Shi:2021hwx}. The obtained nuclear suppression factors for charged hadron and dijet production  are shown in Figs.~\ref{fig:rpa_sinc} and~\ref{fig:rea_jet}.

This rapid progress towards the NLO accuracy has brought the field to the point where precision level studies of saturation phenomena are becoming feasible. As saturation effects are typically expected to be only moderate~\cite{Armesto:2022mxy}, there is likely no smoking gun for gluon saturation even at the EIC. Instead, it will be crucial to perform global analyses where different DIS and p+A observables are simultaneously included. As there is always dependence on the non-perturbative input to the small-$x$ evolution equation, in such analyses it will also be crucial to properly take into account uncertainties in this non-perturbative input (and in other non-perturbative ingredients such as in the vector meson wave function when calculating exclusive vector meson production). At the moment this non-perturbative input does not have any uncertainty estimates available at NLO~\cite{Beuf:2020dxl}, but first steps to include uncertainties in the extraction of the BK evolution initial condition~\cite{Casuga:2023dcf} and the fluctuating proton geometry~\cite{Mantysaari:2022ffw} at leading order have been taken recently.

\section{Longitudinal dynamics}

\begin{figure}[tb]
 \begin{minipage}{.49\textwidth}
  \centering
  \includegraphics[width=\textwidth]{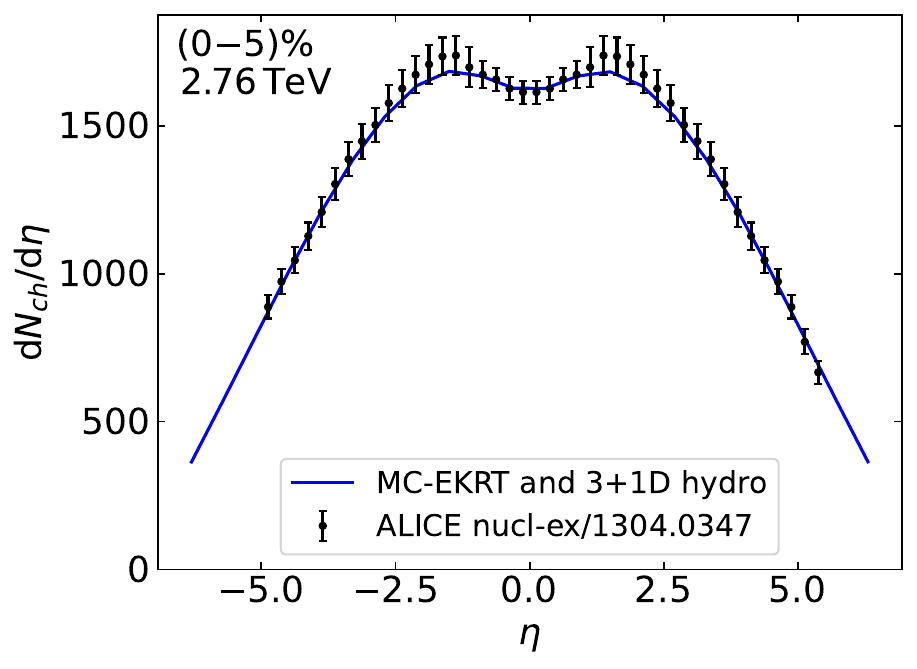}
  \caption{Pseudorapidity distribution of charged hadron multiplicity calculated from the 3D EKRT model~\cite{Kuha} compared to the ALICE data.}
  \label{fig:ekrt}
 \end{minipage}
 \quad 
 \begin{minipage}{.49\textwidth}
 \centering 
\centering\includegraphics[width=0.8\textwidth]{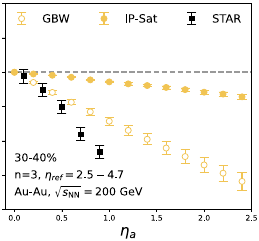}
        \begin{tikzpicture}[overlay]
         \node[anchor=south west,rotate=90] at (-5.5cm,1.7cm) {$r_n(\eta_a,\eta_b)$ }; 
         \node[anchor=south west,rotate=90] at (-5.1cm,0.8cm) {\small $0.85$};
         \node[anchor=south west,rotate=90] at (-5.1cm,2.2cm) {\small $0.95$};
         \node[anchor=south west,rotate=90] at (-5.1cm,3.6cm) {\small $1.05$};
         
    \end{tikzpicture}
\caption{Flow decorrelation estimated form the initial spatial eccentricity in the McDipper model compared to the STAR data~\cite{Nie:2019bgd}. Figure adopted from Ref.~\cite{Garcia-Montero:2023gex}.}
\label{fig:mcdipper}       
\end{minipage}

\end{figure}

Many state-of-the-art descriptions of QGP evolution use 3+1D hydrodynamical simulations and hadronic afterburners. 
In order to fully describe longitudinal dynamics in heavy ion collisions, a realistic $x$-dependent initial condition is also necessary.  
This energy dependence has been included for example in the recent \trento-3D~\cite{Soeder:2023vdn} initial state parametrization.

In weak coupling approach one can again apply either collinear factorization or CGC to go beyond midrapidity. In the  EKRT model~\cite{Paatelainen:2013eea} the input is an $x$-dependent nuclear PDF such as EPPS21~\cite{Eskola:2021nhw}. In this conference, recent developments to the EKRT model  were presented~\cite{Kuha}, including spatially dependent nuclear PDFs with event-by-event fluctuations, a dynamical event-by-event saturation criterium based on minijet production, minijet multiplicity fluctuations and global energy conservation. When this new 3D initial state description is coupled to 3+1D relativistic hydrodynamics, a good description of key heavy ion observables away from midrapidity is obtained as illustrated in Fig.~\ref{fig:ekrt}.

In CGC approach new developments towards a 3D initial condition were also presented. In the new McDipper initial condition~\cite{Garcia-Montero:2023gex} the initial quark and gluon production is calculated in a similar manner as inclusive particle production discussed in Sec.~\ref{sec:saturation}, with the proton small-$x$ structure being described by a parametrization fitted to HERA data. Although currently  the parton production is computed  at leading order accuracy, extensions to higher order accuracy are in principle possible. Using  initial state estimators, a good description of key heavy ion observables is obtained, although the flow decorrelation is  underestimated as shown in Fig.~\ref{fig:mcdipper}.

The $x$-dependence can also be calculated perturbatively by solving the JIMWLK equation. In this conference, the initial state geometry and momentum correlations obtained from the JIMWLK evolution in Ref.~\cite{Schenke:2022mjv} were presented. These results suggest that the initial momentum correlations are short-range in rapidity, unlike the event geometry for which correlations vanish much more slowly when the rapidity separation increases. The longitudinal structure of nuclei governed by the JIMWLK evolution can also be coupled to 3D classical Yang-Mills simulations to determine the time evolution after the collision before QGP is formed. First such implementation was recently shown in Ref.~\cite{McDonald:2023qwc}. When coupled to 3+1D hydrodynamical simulations, a good description of particle spectra, mean $p_T$ and flow harmonics at midrapidity is obtained, but again not enough longitudinal decorrelation for flow is obtained. This can be seen to suggest a need for an additional source of fluctuations.

\section{Conclusions}

The initial state of heavy ion collisions contains a vast amount of interesting fundamental physics, and a realistic initial state description is also crucial to probe in detail the QGP properties. At the moment there is a rapid development to include a realistic description for longitudinal dynamics which enables one to compare simulations of heavy ion collisions to observables away from midrapidity, and to understand the saturation effects at precision level.

The initial state can be inferred directly from heavy ion collisions, or probed in other scattering processes such as deep inelastic scattering or proton-nucleus collisions. A simultaneous description of heavy ion initial state and other collider data can be obtained in collinear factorization based approaches such as EKRT, or in Color Glass Condensate based implementations such as IP-Glasma. For EKRT, recent developments to include longitudinal dynamics was presented in this conference. In CGC based initial state descriptions, there is currently rapid progress in the field to include higher order corrections. These developments are crucial to both enable accurate studies of gluon saturation phenomena especially in the next decade when the Electron-Ion Collider~\cite{AbdulKhalek:2021gbh} becomes operational, and to develop precise initial state descriptions for heavy ion collisions including a perturbatively calculated $x$-dependence. 

\mbox{}\\

\noindent \emph{Acknowledgements}\\
H.M. is supported by the Research Council of Finland, the Centre of Excellence in Quark Matter, and projects 338263 and 346567, and under the European Union’s Horizon 2020 research and innovation programme by the European Research Council (ERC, grant agreement No. ERC-2018-ADG-835105 YoctoLHC) and by the STRONG-2020 project (grant agreement No. 824093).

\bibliography{refs}

\end{document}